\documentclass{WileyMSP-template}
\usepackage{float}
\usepackage{caption, threeparttable}

\usepackage{amsmath}
\usepackage{amssymb}
\usepackage{xcolor}
\usepackage{graphicx}
\usepackage{graphicx}
\usepackage{dcolumn}
\usepackage{bm}

\usepackage{graphicx}
\usepackage{dcolumn}
\usepackage{bm}
\usepackage{amsmath}
\usepackage{xcolor} 
\usepackage{hyperref}
\usepackage{graphicx}
\usepackage{physics}
\usepackage{mathtools}
\usepackage{dsfont}
\usepackage[superscript,biblabel]{cite}
\makeatletter \renewcommand{\@citess}[1]{\textsuperscript{\,[#1]}} \makeatother
\usepackage{soul} 

\begin{document}

\title{Unveiling the link between quantum ghost imaging and Grover's quantum searching algorithm}

\maketitle


\author{Neelan Gounden}
\author{Fazilah Nothlawala}
\author{Paola C. Obando}
\author{Thomas Konrad}
\author{Andrew Forbes}
and \author{Isaac Nape*}

\begin{affiliations}
Neelan Gounden, Fazilah Nothlawala, Paola C. Obando, Andrew Forbes \\

Structured Light Laboratory, School of Physics, University of the Witwatersrand, Johannesburg, 2000, South Africa

Thomas Konrad\\
School of Chemistry and Physics, University of KwaZulu-Natal, Private Bag X54001, Durban 4000, South Africa

Isaac Nape\\
Structured Light Laboratory, School of Physics, University of the Witwatersrand, Johannesburg, 2000, South Africa\\
isaac.nape@wits.ac.za
\end{affiliations}



\begin{abstract}

Photonic quantum technologies have become pivotal in the implementation of communication, imaging and computing modalities. Among these applications, quantum ghost imaging (GI) exploits photon correlations to surpass classical limits, with recent advances in spatial-mode encoding and phase imaging. In parallel, all-optical computing offers powerful, passive-light processing capabilities. Here, we explore the intersection of these domains, revealing a conceptual and operational link between GI and Grover’s quantum search algorithm (GSA) which is designed to search for elements in an unstructured database. Here, the elements are encoded as phases in the position basis states of photons.  To show this, we use entangled photon pairs, with one photon encoding the oracle features while the other photon is used to find the marked element.
\end{abstract}


\section*{Introduction}

Photonics continues to play a pivotal role in advancing quantum technologies, enabling platforms that outperform their classical counterparts in a range of applications. Quantum photonics, in particular, harnesses non-classical properties of light, such as entanglement, superposition, and wave-particle duality, to unlock secure quantum communication \cite{forbes2024quantum, yin2020entanglement, liao2017satellite}, energy-efficient quantum computation \cite{o2007optical, zhong2020quantum, ladd2010quantum, nimbe2021models, riste2017demonstration}, and high-resolution, low-photon-count imaging modalities \cite{gilaberte2019perspectives, moreau2019imaging, moodley2023all, gregory2020imaging}. These advances mark a shift toward practical quantum information systems powered by light.\\
A particularly promising frontier lies in quantum imaging, where quantum correlations are used to surpass classical imaging limits. For example, quantum ghost imaging (GI) leverages entangled or correlated photons to image objects using photons that never directly interact with the object \cite{Moreau18, Pittman95, Abouraddy2001, erkmen2010ghost}. Originally seen as a quantum-inspired curiosity, GI has matured into a versatile tool for high-resolution biological imaging \cite{zhang2024quantum}, ultrafast temporal measurements \cite{ryczkowski2016ghost}, and fundamental tests of quantum mechanics \cite{jack2009holographic, aspden2013epr}. These developments, now thoroughly reviewed \cite{shih2007quantum, defienne2024advances, padgett2017introduction, erkmen2010ghost, moodley2024advances}, highlight the growing relevance of quantum imaging. Implementations now span from thermal light \cite{ferri2005high, bennink2002two} to entangled photon sources via SPDC \cite{moreau2019imaging}, incorporating sophisticated states such as coalesced two-photon and entanglement-swapped channels \cite{bornman2019ghost}.\\
GI has also evolved in terms of information encoding, with spatial-mode-based implementations using Hadamard-Walsh \cite{wang2016fast}, Fourier \cite{zhang2015single}, and other modal bases. Notably, Hadamard-based imaging has enabled quantum phase imaging \cite{sephton2023revealing}, allowing phase-only objects to be imaged through intensity measurements. Parallel to these advances, all-optical computing is gaining prominence, where light is processed using purely passive optical components, often guided by inverse design principles. Such systems have successfully demonstrated the classification of optical digits, structured light patterns, and even the synthesis of complex unitary transformations. This convergence of quantum imaging and optical computing is driving a new era of photonic information processing, merging the Fourier optics formalism with principles from machine learning and quantum algorithms.\\
In addition to this fusion of imaging and computing, quantum computing algorithms themselves are being translated into the optical domain. Grover’s search algorithm (GSA), among the most celebrated quantum algorithms, has been demonstrated using polarization encoding \cite{kwiat2000grover}, orbital angular momentum modes \cite{perez2018first}, and more recently in photonic integrated circuits \cite{tabia2016recursive}. Interestingly, the operations used in GSA, particularly its phase inversion and diffusion steps, bear a strong resemblance to the phase manipulation techniques employed in phase-contrast microscopy, suggesting a deeper link between quantum computing and quantum imaging.\\
In this work, we demonstrate for the first time a link between GSA and GI. We show that both processes rely on analogous phase operations and measurement protocols, effectively unifying concepts from imaging and computing. To illustrate this, we use photon pairs generated from SPDC and show that the pixel states in the GI protocol can be used as a high-dimensional basis where one photon encodes information from the oracle, the other is used to search the database, exploiting the spatial entanglement between the photons.

\begin{figure*}[t]
		\centering
		\includegraphics[width=1\linewidth]{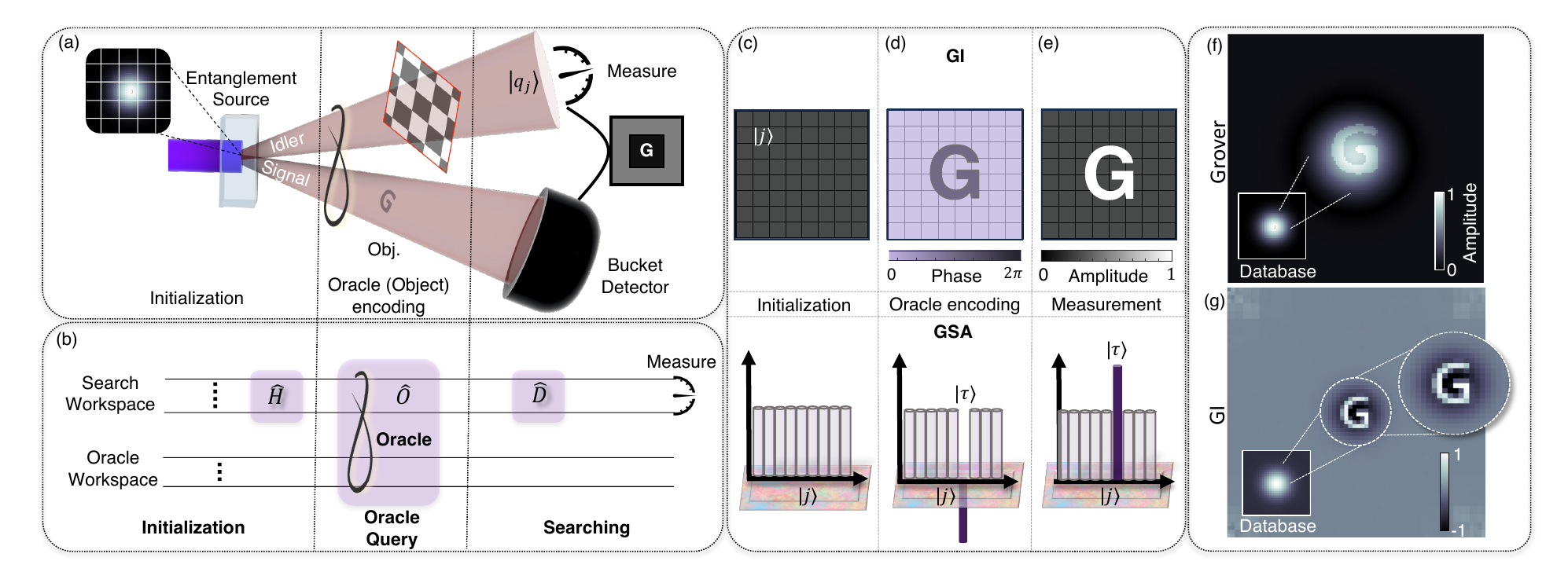}
  \caption{\textbf{Quantum ghost imaging (GI) as an analogue of Grover's search algorithm (GSA)}. 
 (\textbf{a}) Conceptual sketch of a GI  setup. High dimensional entangled bi-photons are generated via spontaneous parametric down conversion (SPDC) using a non-linear crystal (NC) and are separated spatially. One photon, the signal, interacts with a phase (transparent) object, is imbued with spatial information, and then collected by a non-spatially resolving (bucket) detector. Its entangled twin, the idler photon, is measured using projective masks (or states, $\ket{q_j}$) that subsequently map the phases of the object to amplitudes.  The joint measurements (coincidences) are then used to reconstruct an image of the object, producing an image that reveals the phases as amplitudes. (\textbf{b}) A parallel can be drawn between GI and GSA processes. At the  initialization stage, the search database and the oracle workspace are created, analogous to the entangled subsystems in the GI setup.  Subsequently, during the oracle stage, the target elements $\tau$ are marked, analogous to the action of the object in GI. The searching process is performed on the search workspace (equivalent to the idler photon), followed by the application of the diffusion operator $(\hat{D})$ and subsequent measurement in the logical basis.   A graphical outline of the two protocols shows the initialization (\textbf{c}), oracle marking (\textbf{d}) and measurement stages (\textbf{e}) on the respective Hilbert spaces, i.e. corresponding to high dimensional pixel basis states (top panel) and logical basis states constructed from multi-qubit spaces (bottom panel). (\textbf{f}) Simulation of Grover's search algorithm applied to an encoded object. Here the image is reconstructed using a measurement approach where the Grover diffusion operator is absorbed into the measurement basis using a $128\times128$ pixel basis. The inset showing the Gaussian profile maps the active Hilbert space that the oracle operators act on - i.e. the database. In the experiment, it corresponds to the amplitude coefficient ($\lambda_j$) over the pixel states $\ket{j}$.(\textbf{g}) Simulation using the Hadamard adjusted basis $\ket{q_j}$ in the measurement procedure via GI.}
\label{fig:comparison}
\end{figure*}

\vspace{0.25in}

\section*{Theory and Concept}

\textbf{Ghost imaging vs Grover's algorithm}. 
The similarities between GI and GSA can be seen in the schematics shown in Figure \ref{fig:comparison} (a) and  Figure \ref{fig:comparison} (b), respectively. The two protocols have three stages: (i) preparation, (ii) object or oracle  encoding and (iii) the searching and measurement stage. Moreover, both schemes can have two separate subsystems in which the object of interest is encoded and another where the searching operation is performed. For GI, one photon/subsystem from an entangled pair encodes the object's spatial information by marking the photon's spatial amplitude (and phase) while the other particle is used to extract the encoded features. In comparison, GSA makes use of a similar structure \cite{grover96} ; instead of a single particle, a collection of qubits are used to form two registers, being, one for the search space (equivalent to our detection photon), while the other, the oracle workspace, takes up the same role as the photon that interacts with an object. 
Next, we outline these equivalences mathematically. 

\vspace{0.25in}
\textbf{Initialization and element marking.}
The GI protocol makes use of two spatially separated photons, i.e. signal and idler, as a resource acting similarly to the search and oracle workspaces from Grover's search algorithm. In our scheme, they are generated from a nonlinear crystal (see Experimental Setup) through the process of spontaneous parametric down conversion (SPDC)  and  can be described by the nonseparable superposition state:
\begin{align}
\label{eq:entanglingOP}
 \ket{\Psi_{si}} &=  \sum^{M-1}_{j=0} \lambda_j\ket{j}_s\ket{j}_i,
\end{align}
 occupying the combined Hilbert space, $\mathcal{H}_{si}=\mathcal{H}_s\otimes\mathcal{H}_i$, formed from the tensor product of discrete position (or pixel) states $\ket{j}$  \cite{valencia2020high} with $j$ $=$ 0, 1,... $N-1$, for the signal and idler photon, each defined on Hilbert spaces $\mathcal{H}_s$ and $\mathcal{H}_i$, respectively. The dimensions of the Hilbert space for each subsystem is given by $M$ while the Hilbert space within which the oracle operates is given by $N\leq M $. The coefficients, $\lambda_j$, set the size/dimensions ($N$) of the oracle workspace and can be computed from the Schmidt number. Our experiment produced $N = 549 $ dimensions of pixel states that can be used for encoding  information \cite{moreau2018resolution}. Notably, the benefit of using pixel states of photons is that the photons are already defined on higher dimensional state spaces, so that fewer particles are needed to encode multiple elements of the database.
 
\vspace{0.25in}
\textbf{Oracle  encoding.}
We visualise the encoding space for the signal and oracle-workspace in Figure \ref{fig:comparison} (c) represented as pixels (top panel) and in terms of logical basis states (bottom panel), $\ket{j}$, for the GI and GSA, respectively. Next, the signal photon (analogue of the oracle workspace), interacts with an unknown object, labelled as Obj. in the Figure \ref{fig:comparison} (a), producing the state
\begin{equation}\label{eq:objstate}
(\hat{O} \otimes I_M) \ket{\Psi_{si}}= \sum^{M-1}_{j=0} \lambda_j o_{j} \ket{j}_{s} \ket{j}_{i},
\end{equation}
where $\hat{O}=\sum_{j=0}^{M-1}o_j\ket{j}\bra{j}$ is an operator that encodes the elements via the coefficients $o_{j}$ , which represent the entries of the database onto the corresponding pixels $\ket{j}$ on a transversal plane of the signal photon -  the same operation happens in GSA. The coefficients mark the desired pixel states (or database elements) with $\pi$ phases, as shown in the oracle encoding stage of Figure \ref{fig:comparison} (d), where the amplitudes of the marked elements are inverted. Therefore, selective amplitude inversion is performed, where the amplitudes for the target elements ($\ket{\tau}$) are converted from positive to negative amplitudes so that $o_j = -1^{f(j)}$ where ${f(j=\tau)} = 1$ for the marked elements and is set to  $f(j \neq \tau) = 0$  otherwise.

After the element marking, the information encoded in the oracle space and signal are transferred to the search workspace and idler photon. To achieve this in the GI protocol, the signal photon is detected using a bucket detector (See Figure \ref{fig:comparison} (a)), that is, a detector that cannot resolve the encoded spatial information \cite{tasca2013influence}. Here, our bucket detector is a single mode fiber and an avalanche photo-diode detector (See Experimental Setup) that only accepts a Gaussian mode and therefore acts as a projective filter that heralds (marks) and transfers the coefficients $o_j$ from the signal to the idler arm. Similarly, in GSA the conditional operations that are applied by the oracle are used to herald the desired elements in the search workspace. At this stage, the state of the idler, and equivalently the search space, is given by 
\begin{equation}
\ket{\Psi_{i}}=\sum^{M-1}_{j=0} o^{'}_j \ket{j}_{i},
\label{fig: idler-search}
\end{equation}
with $ o^{'}_j \equiv  \lambda_j o_{j}$. Now we remain with the idler photon in GI and likewise the search workspace in GSA. At this point, both protocols execute a search procedure in order to recover the information about the marked elements.

\vspace{0.25in}
\textbf{Searching for the marked elements}.
Now that we have shown that the GI and GSA are able to execute the element marking steps in a similar fashion, all that remains is to demonstrate the searching procedures. In Figure \ref{fig:comparison} (e) we show the resulting outcomes for GI and GSA,  each showing that the marked elements are mapped from the inverted amplitudes to positive amplitudes, with the marked elements having higher amplitudes than the unmarked elements. 

In GSA, the amplification of the marked elements is performed by the operator
\begin{align}
\label{Eq:difussorop}
    \hat{D}&= 2\ket{h_0}\bra{h_0}_i - \mathbb{I}_{M},
\end{align}
where $\ket{h_0} = \frac{1}{\sqrt{M}} \sum^{M-1}_{j=0} \ket{j}$ is the uniform superposition state in the logical basis and $\mathbb{I}_M$ is the $M$ dimensional identity matrix. This operation results in the detection probabilities 
\begin{align}
     p_j = | \langle h_0 |\langle j |  \hat{O}  \otimes   \hat{D}     | \Psi_{si} \rangle|^2  &\propto |\langle j | \hat{D} | \Psi_{i} \rangle|^2 \nonumber \\
&= |2 \langle  o^{'} \rangle - o^{'}_j|^2,
     \label{eq:probabilities_Grover}
\end{align}
for each state, $\ket{j}$, in the database.  The probability amplitudes, $2\langle o^{'} \rangle - o^{'}_j $,  represent a reflection of the amplitude, $o_j$, about the mean of $|\Psi_i\rangle$ with $ \langle o^{'} \rangle \equiv \langle h_0 |\Psi_j\rangle = \sum_{j} \lambda_j o_j /M$, representing the mean of the probability amplitudes of the coefficients of the search space state $|\Psi_i \rangle$. As shown in Figure \ref{fig:comparison} (e),  the marked element's probability is enhanced and is higher than that of the unmarked elements (due to the reflection of the marked probabilities about the mean probability of the search space). However, the unmarked elements still have nonzero probabilities. This is illustrated in the simulated reconstruction using GSA in Figure \ref{fig:comparison} (f), where the marked elements represented by the 'G' are enhanced amplitudes, however the unmarked elements around the 'G' still have non-zero amplitudes.

The impact of the unmarked elements can be seen if we recast GSA in terms of GI;   instead of applying the diffusion operator to the state, $\ket{\Psi_i}$,  we can absorb it into the measurement basis states, $\ket{j}$, resulting in the unnormalised measurement states, 
\begin{equation}
    \hat{D} \ket{j} \propto \frac{2}{\sqrt{M}}\ket{h_0} - \ket{j},
    \label{eq:groverproj}
\end{equation}
where each basis element is superimposed with the uniform superposition state, $\ket{h_0}$.  In this perspective, Equation \ref{eq:probabilities_Grover} measures the probability of detecting the marked element in the idler (search workspace)  given that it has been marked in the signal (oracle workspace) photon.  As a result, we can visualise the results  as an image by mapping each pixel state, $\ket{j}$,  onto pixels  $j$ as can be seen in Figure \ref{fig:comparison} (g).  In the example, the marked elements are modulated by a background Gaussian amplitude, encoded by the coefficients $\lambda_j$, which determine the initial  amplitudes of each database element before the marking and amplification step $\lambda_j = \frac{1}{\sqrt{N}}$, for all $j$ , with each pixel state having an equal contribution.   In GI this is determined by the entanglement source and in our case is a Gaussian distribution over the pixel states. 
It is important to note the difference between GSA and GI: 
In GSA, a single Grover iteration can be represented as a product of $\hat{O}\cdot \hat{D}$ (consecutive application of oracle and diffusion operations on the state). In contrast, in GI,  it is a tensor product $\hat{O} \otimes \hat{D}$ (i.e. a simultaneous application of the oracle and diffusion operations on the state). The equivalence of consecutive operations on a single system and simultaneous operations on a pair of entangled systems is due to the Choi Jamiolkowski isomorphism between operations and tensor products of states.

\vspace{0.25in}
\textbf{Searching for marked elements with an adjusted basis from GI}

\begin{figure*}[h!]
	\centering
	\includegraphics[width=1\linewidth]{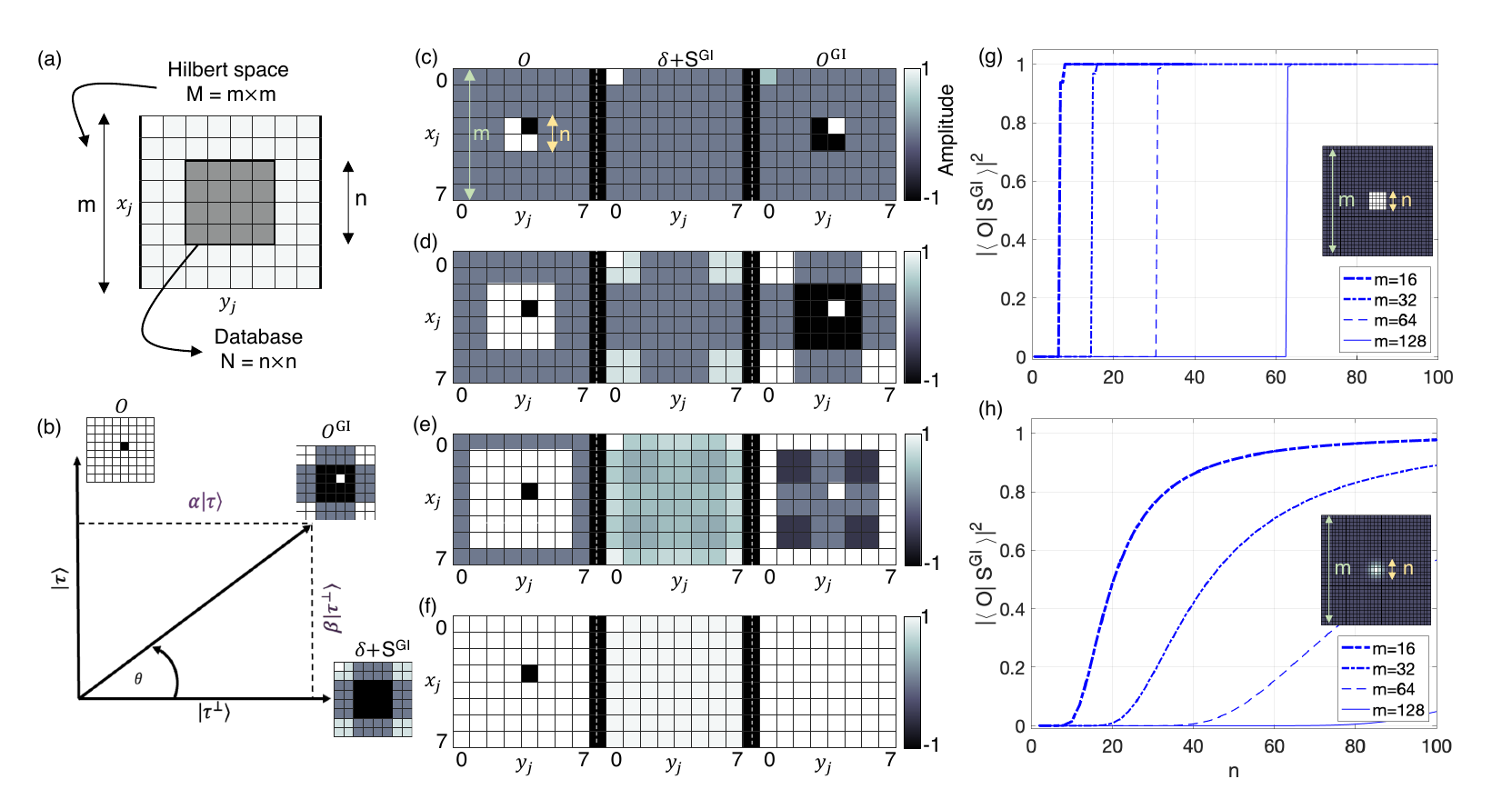}
  \caption{
\textbf{Partitioning of the solution and non-solution Hilbert spaces}. (\textbf{a}) The neighbourhood of pixel states, $N = n \times n$ (the database), is embedded within the larger set of pixels, $M = m \times m$. (\textbf{b}) The computed ghost image, $O^{\text{GI}}$, is decomposed into orthogonal complements, $\ket{\tau}$ and $\ket{\tau^\perp}$ . The solution space $\ket{\tau}$ corresponds to the marked elements of $O$, while $\ket{\tau^\perp}$ includes the unmarked elements and contributions from the remaining pixels outside  the $N = n \times n$ pixels that include contributions of $\delta_0+ S^{\text{GI}}$. (\textbf{c})-(\textbf{f}), illustration of the separation of $\ket{\tau}$ and $\ket{\tau^\perp}$ for $m=8$ and various $n$ values (2, 4, 6, 8). As $n$ increases, the overlap between solution and non-solution spaces becomes non-zero when $n > m/2$. (\textbf{g}) Illustrates numerical simulations indicating the effect in the overlap $\langle O|S^{GI}\rangle$ for varying $m$ and $n$ values, the database is constructed as a collection of pixels with an even distribution. (\textbf{h}) Illustrates numerical simulations indicating the effect in the overlap $\langle O|S^{GI}\rangle$ for varying $m$ and $n$ values, the database is constructed as a gaussian distribution over pixel states.}
    \label{fig:solution_nonsoliution}
	\end{figure*}
\textbf{Unmasking the marked elements with alternative GI projections.}
The single pixel measurement approach that emerges when we absorb the Grover operator into the detection basis is similar to single pixel GI, where one pixel is measured at a time.  Here, we show that the basis can be adjusted to project multiple pixels simultaneously while harnessing the GI protocol's ability to perform phase measurements  \cite{sephton2023revealing}. This will make it possible to detect the elements that have been marked without additional amplification procedures. Therefore, we will show that performing traditional GI can indeed suppress the contribution of the unmarked elements in a single step but by inverting them in the reconstruction of the object features as shown in Figure \ref{fig:comparison} (g).

To achieve this,  we use the superposition states
\begin{equation}
\ket{q_j}  = \frac{ \ket{h_0} - \ket{h_j}}{\sqrt{2}},
\end{equation}
as our measurement basis. Here, $\ket{h_j} = \hat{H} \ket{j}$  are the Hadamard basis states that can be obtained by applying the Hadamard transform to the logical basis  $\ket{j}$ and  $\ket{h_0} = \hat{H} \ket{0}$ is the uniform superpositions state that can be obtained by applying the Hadamard gate to the first basis element, $\ket{0}$.  Accordingly,  each basis mode, $\ket{h_j}$, is superimposed with the reference state $\ket{h_0}$, unlike the single pixel state. Consequently, the detection probability, given the idler photon  after the oracle marking stage becomes,
\begin{align}
    p_j &= | \langle q_j | \Psi_{i}  \rangle |^2, \nonumber\\
     &= | \langle  o^{'} \rangle - \tilde{o}_j  |^2,
    \label{eq:GHostprob}
\end{align}
where $ \tilde{o}_j = \sum_{k=0}^{M-1} \lambda_k o_k \langle h_j| k \rangle = \sum_{k=0}^{M-1} \lambda_k o_k h_{jk}$ and $h_{jk}$ are the components of the $j^{\text{th}}$ state while $ \langle o^{'} \rangle \equiv\langle h_0 |\Psi_j\rangle = \sum_{j} \lambda_j o_j /M$.  In this case, the  probability outcomes measure the fluctuations of the amplitudes about the mean of $|\Psi_i \rangle $. 

Next, we can compute the reconstructed image from  
\begin{equation}
O^{\text{GI}}(x, y) =  \sum_j p_j \ h_j(x, y),
\label{eq:imreconhad}
\end{equation}
using the transmission masks $  \ket{h_j} \rightarrow h_j(x, y)$ for the $j^{\text{th}}$ state  depicted in Figure ~\ref{fig:comparison} (c). Figure \ref{fig:comparison} (g) shows a simulation example of such an image which is obtained from measuring the probabilities $p_j$.  

\vspace{0.25in}
It can be shown (from Equation \ref{eq:GHostprob} and Equation \ref{eq:imreconhad}) that the image is given by
\begin{align}
O^{GI}(x, y) &=   |\langle o^{'} \rangle_M|^2 \sqrt{M}  \delta_{j0}(x_0, y_0) + S^{GI}(x, y)  \nonumber  \\ & \  \ \  - |\langle o^{'} \rangle_M| \cdot O (x, y) ,
\label{eq:ghostIM}
\end{align}
where $\delta_{j0}(x_0, y_0) =\frac{1}{\sqrt{M}} \sum_{j}  h_j(x, y) $ maps onto the first pixel state ($\ket{j =0}$) such that only the first pixel has nonzero amplitude,  and $S^{GI}(x, y) = \sum_{j}(\tilde{o}_j)^2 h_j(x, y)$  is the Hadamard transform of the power spectrum of the oracle components $\lambda_{j} o{_j}$ in the Hadamard basis while $O (x, y) = \sum^{M-1}_{j=0} \tilde{o}_j h_j(x, y)$ is the information encoded by the oracle, $O (x, y)$.  We can decompose the reconstructed image into two components, i.e. $O^{GI}= O_{\tau} + O_{\tau^{\perp}} $, the part that contains the solution ($O_{\tau}$) and the part that doesn't ($O_{\tau^{\perp}} = \delta_{j0} + S^{\text{GI}}$).

Next, we establish the conditions under which these subspaces are independent, so that the marked elements are detectable after the imaging procedure is implemented.

\vspace{0.25in}
\textbf{Separating the marked and unmarked elements.}
The negative sign in front of the term containing the $O(\cdot)$ factor in Equation \ref{eq:ghostIM} ensures that the unmarked elements have negative amplitudes after the image reconstruction, while the marked elements have positive amplitudes as seen in the examples presented in Figure \ref{fig:examples}. However, the inversion depends on the relative size of the total Hilbert space ($M$) over which the measurements are performed and the size of the Hilbert space ($N<M$) that contains the database elements. The partitions of these Hilbert spaces are shown graphically in  Figure \ref{fig:solution_nonsoliution} (a). This is because the SPDC illuminates part of the Hilbert space while the measurements are performed over a larger Hilbert space.  

To illustrate this point,   first we define the neighbourhood of  pixel states, $N = n \times n$ that are embedded in the $M = m\times m$ set of pixels as shown in Figure \ref{fig:solution_nonsoliution} (a).   The computed GI image, $O^{\text{GI}}$, can now be decomposed into  orthogonal complements,  $\ket{\tau}$ and $\ket{\tau^\perp}$, shown in Figure \ref{fig:solution_nonsoliution} (b), containing the solution space corresponding to the marked elements of $O$ and non-solution spaces corresponding to the unmarked elements in $O$ and the images $\delta + S^{\text{GI}}$, respectively.  As shown using the two-dimensional axis,  $\ket{\tau}$, labels the marked element contained on the horizontal axis of Figure \ref{fig:solution_nonsoliution} (b)  whereas $\ket{\tau^\perp}$ is composed of the inverted elements and every other pixel states other than the ones that were marked. The diagonal axis shows that the marked elements of  $O(\cdot)$ can be distinguished from the unmarked elements and contributions from  $S^{GI}(\cdot)$  and  $\delta_{0}(\cdot)$ (containing pixel states outside the domain of  $N= n\times n$ pixels).  In a conventional GSA, $\hat{G}_D$, rotates the space by $\theta$ radians per iteration, rotating the state vector closer to $\ket{\tau}$, therefore amplifying the solution space \cite{nielsen_chuang_2010}. In our case, the procedure separates the solution and non-solution spaces by virtue of the amplitude inversion and non-overlapping nature of the other subspaces ($\delta + S^{\text{GI}}$) that are outside the $n \times n $ neighbourhood.

\vspace{0.25in}

Next, we show that the separation of the  $\ket{\tau}$ and $\ket{\tau^\perp}$ states in the pixel space, depends on the size of the database.   We make this comparison for $m = 8$  for various dimensions of the database given by  $n \ =$  2, 4, 6 and 8 in Figure \ref{fig:solution_nonsoliution} (c), (d), (e) and (f) respectively. For Figure \ref{fig:solution_nonsoliution} (c) and (d),  we see that  $\delta + S^{GI}$ does not overlap with the $n\times n$  database. The final image $O^{\text{GI}} =   \delta+S^{\text{GI}}  - O$ confirms this. Moreover images Figure \ref{fig:solution_nonsoliution} (c) and (d) which satisfy $n \leq m/2$, both have no overlap between the $O$ and $S^{GI}$ subspaces. However, from Figure \ref{fig:solution_nonsoliution} (e)  and (f) we see that the overlap between the $O$ and $S^{GI}$ subspaces increases as n increases.. This is the case for $n>m/2$ which will produce a non-zero overlap between the two subspaces.

\vspace{0.25in}

For larger scales this effect can also be observed.
In Figure \ref{fig:solution_nonsoliution} (g), it can be shown for a given Hilbert space size ($M$),the overlap between the solution and non-solution subspaces varies when the database size ($N$) changes. There is no overlap between the subspaces $O$ and $S^{GI}$ when $n\leq m/2$. Beyond the point where $n>m/2$ there is a complete overlap between the subspaces $O$ and $S^{GI}$. 
Experimentally, the database size is determined by the profile of the beam on the Hilbert space (a spatial light modulator screen). Since the beam profile is gaussian, Figure \ref{fig:solution_nonsoliution} (h) represents a numerical simulation varying $n\approx 2\sqrt{\omega}$ ,where $\omega$ represents the second moment of the gaussian beam. Similar effects are observed in Figure \ref{fig:solution_nonsoliution} (h) ,as $n$ changes. The gaussian profile of the database has a radially varying intensity when being compared to the even intensity of the database in Figure \ref{fig:solution_nonsoliution} (g), this leads to a gradual increase in the overlap between the subspaces $O$ and $S^{GI}$ as $n$ increases. When the size of the gaussian beam is sufficiently large the overlap between the two subspaces will converge to 1.
 
\vspace{0.25in}



\section*{Experimental setup}
\begin{figure*}[h!]
		\centering
		\includegraphics[width=0.95\linewidth]{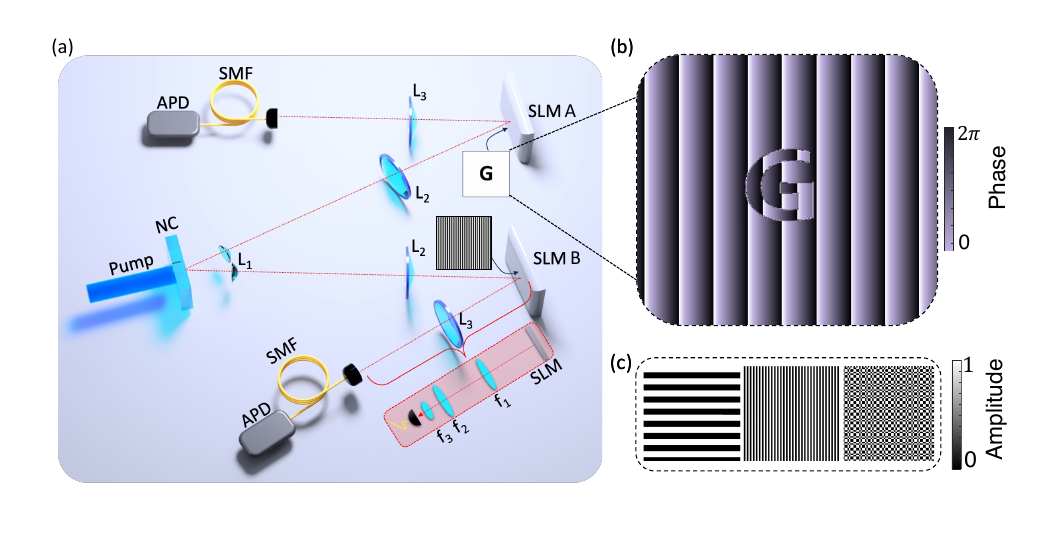}
		\caption{\textbf{Experimental object reconstruction process}. (\textbf{a}) Shows a schematic diagram of the implemented quantum searching setup. A laser produces a pump photon, with a wavelength of $404.2$ nm, which is incident on a non-linear crystal (NC) which leads to the generation of a pair of entangled photons each possessing a $808.4$ nm wavelength. The entangled photons are spatially separated and imaged onto two spatial light modulators (SLMs), photon A is incident on SLM A which displays the object while photon B is incident on SLM B which displays the projective masks. Subsequently, the photons are collected by coupling each photon into a single mode fiber (SMF), which is connected to an avalanche photo-detector (APD). The photons are detected in coincidence and counted by a coincidence counter. The coincidences are used to reconstruct an image of the object.  (\textbf{b}) Shows an example of an object being displayed on the SLM A (with a phase grating). (\textbf{c}) Illustrates examples of Hadamard projective masks used to spatially resolve photon B.}
		\label{fig:Experimentalsetup}
	\end{figure*}

\noindent An ultraviolet pump photon of wavelength ($\lambda = 404.2$ nm) with OAM $\ell\hbar = 0$ was incident on a PPKTP non-linear crystal (labelled NC in Figure ~\ref{fig:Experimentalsetup} (a)), of length $2$ mm, to produce two photons of wavelength ($\lambda = 808.4$ nm) by spontaneous parametric downconversion (SPDC). The beam's radius out of the crystal was $0.3$ mm. The collinearity of the SPDC beam (or specifically the pair of down-converted photons) was controlled by adjusting the temperature of the crystal. In order to generate a non-collinear SPDC profile and for alignment purposes, the crystal was set to a temperature of $ 57^\circ C $ using an oven. The resolution of the system can be estimated to be approximately 549 pixels \cite{moreau2018resolution}. 
The two photons, signal (photon A) and idler (photon B) were each imaged and magnified with two lenses ($L_1$ and $L_2$) from the crystal plane to the plane of the spatial light modulators (SLM A and SLM B respectively- Holoeye Pluto 2.1). The magnified beam radius of the SPDC was $3$ mm on the SLM screens. Each of the SLM screens consists of $960 \times 960$ pixels with each pixel having dimensions $8 \times 8$ $\mu$m$^2$. SLM A has a digital version of the object ('G') encoded onto its screen, while SLM B has the projective masks encoded onto its screen (superposition of hadamard masks), with example holograms for each shown as insets in Figure \ref{fig:Experimentalsetup} (a). In addition, a phase grating is applied to the projective masks and the object such that the modulated first-order of the diffracted SPDC is incident on the SMFs. The phase grating which is applied can be seen on the hologram in Figure ~\ref{fig:Experimentalsetup} (b). The first-order light which was being modulated off the SLM was demagnified by a factor of 10 using a telescope, the demagnified beam was then Fourier transformed by a 2 mm lens onto the single mode fiber consisting of a fibre core width of 4.4 $\mu$m. The telescope system (constructed with lenses of focal lengths $f_1$ and $f_2$) and the Fourier lens (lens of focal length $f_3$) is represented as $L_3$ as shown in Figure \ref{fig:Experimentalsetup} (a). The SMFs are connected to independent Avalanche-Photo Diodes (APDs),which are used to detect single photon events. These detectors were connected to a coincidence counter (Swabian: Time Tagger) which was used in conjunction with the APDs. The detectors and counter can be used to measure coincidence events obtained from a pair of entangled photons with a gating time of 3 ns and an integration time of 2 s. 

\vspace{0.25in}

\section*{Image reconstruction and processing}

\begin{figure}[h!]
    \centering
    \includegraphics[width=0.6\linewidth]{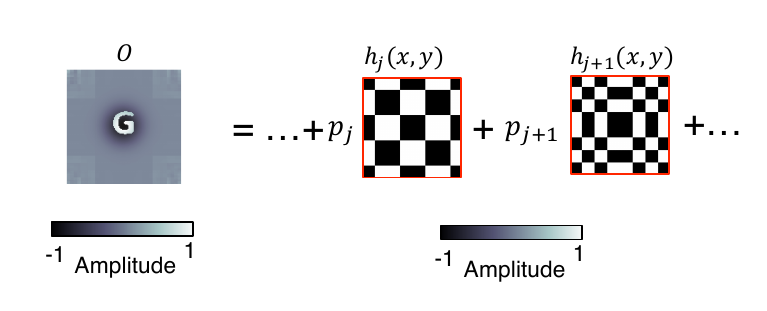}
    \caption{\textbf{Mask synthesis and image reconstruction.} Illustration of the image reconstruction procedure using the Hadamard basis masks $h_j(x, y)$ and the measured probabilities $p_j$. The object $O$ can be reconstructed as a linear superposition of Hadamard basis masks $h_j(x, y)$ with corresponding weighting coefficients $p_j$.}
    \label{fig:imrecon}
\end{figure}


While absorbing the diffusion operator shows that GSA can be performed using GI,  this however only encodes the first iteration of GSA.  For the full implementation, 
$\sqrt{M}$  iterations of the diffusion operator and oracle calls \cite{grover97Q} in required, so that the unmarked elements are suppressed. In our image analogue, applying these iterations can increase the contrast of the image in Figure \ref{fig:comparison} (f).   Furthermore, because the states in Equation \ref{eq:groverproj}, project onto single pixel states $\ket{j}$ with a constant background of pixels given by $\ket{h_0}$, the single pixels can result in noisy measurements if the pixels are small compared to the photon field or if the signal fluctuations are higher than the signal that can be measured by a single pixel.  Next, we show how the measurement basis can be adjusted to account. 

\vspace{0.25in}

The object will obtain a $\pi$ phase shift compared to the surrounding region. The masks are constructed using the Walsh-Hadamard basis. The object is encoded on SLM A. In order to reconstruct the object the basis masks need to be generated, 
using the Walsh functions in one $W_u(x)$, where $u$ indexes the $u^{\text{th}}$  function ranging from 0 to $m-1$ and determines the elements within the function, the 2D Walsh-Hadamard basis states can be constructed by taking the outer product of two one-dimensional Hadamard functions \cite{rodriguez2020towards}:
\begin{align}
h_{j}(x,y) &\in \{  W_u(x) \otimes W_v(y); u, v = 0,.., \sqrt{M}-1\}, 
\end{align}
The projective masks used can be obtained as follows:
\begin{equation}
q_{j}(x,y)=\frac{h_{0}(x,y)-h_{j}(x,y)}{\sqrt{2}},
\end{equation}

where $j \in [0,(m \times m) -1]$. Using these parameters this will lead to $M =m^2$ projective masks. An example of one of these masks are depicted in Figure \ref{fig:imrecon} as $h_j(x,y)$. Since the dimensions of the masks and that of the projective basis differ, super-pixels (pixels consisting of multiple individual SLM pixels) will be used to enable the projective mask to fill the entirety of the SLM screen. The dimensions of these super-pixels can be calculated to be: $d_{super-pixel(32)}=\frac{960}{32}=30$ while $d_{super-pixel(64)}= 15$ and $d_{super-pixel(128)} = 8$.

\vspace{0.25in}

In order to obtain a reconstructed image of the object, we require a linear combination of the projective masks using Equation \ref{eq:imreconhad}
where $p_j$ (Equation \ref{eq:GHostprob}) represents the coefficients for the corresponding projective mask $q_{j}$.
The required coefficients ($p_{j}$) are measured using the setup, as the object is being displayed on SLM A, a projective mask is displayed on SLM B and bi-photon events (coincidences) are recorded over a period of 2 s and a gating time of 3 ns is used. The projective masks are iterated through on SLM B so for the $j$th mask $q_{j}$ the corresponding coefficient $p_j$ can be obtained from the measurement of the normalised coincidences. The single photon events at each detector (A and B) is recorded as well, this allows us to subtract any classical noise (accidentals) which would have contributed to the coincidence detection counts. 

\section*{Results and discussion}

\begin{figure*}[h!]
\includegraphics[width=0.9\linewidth]{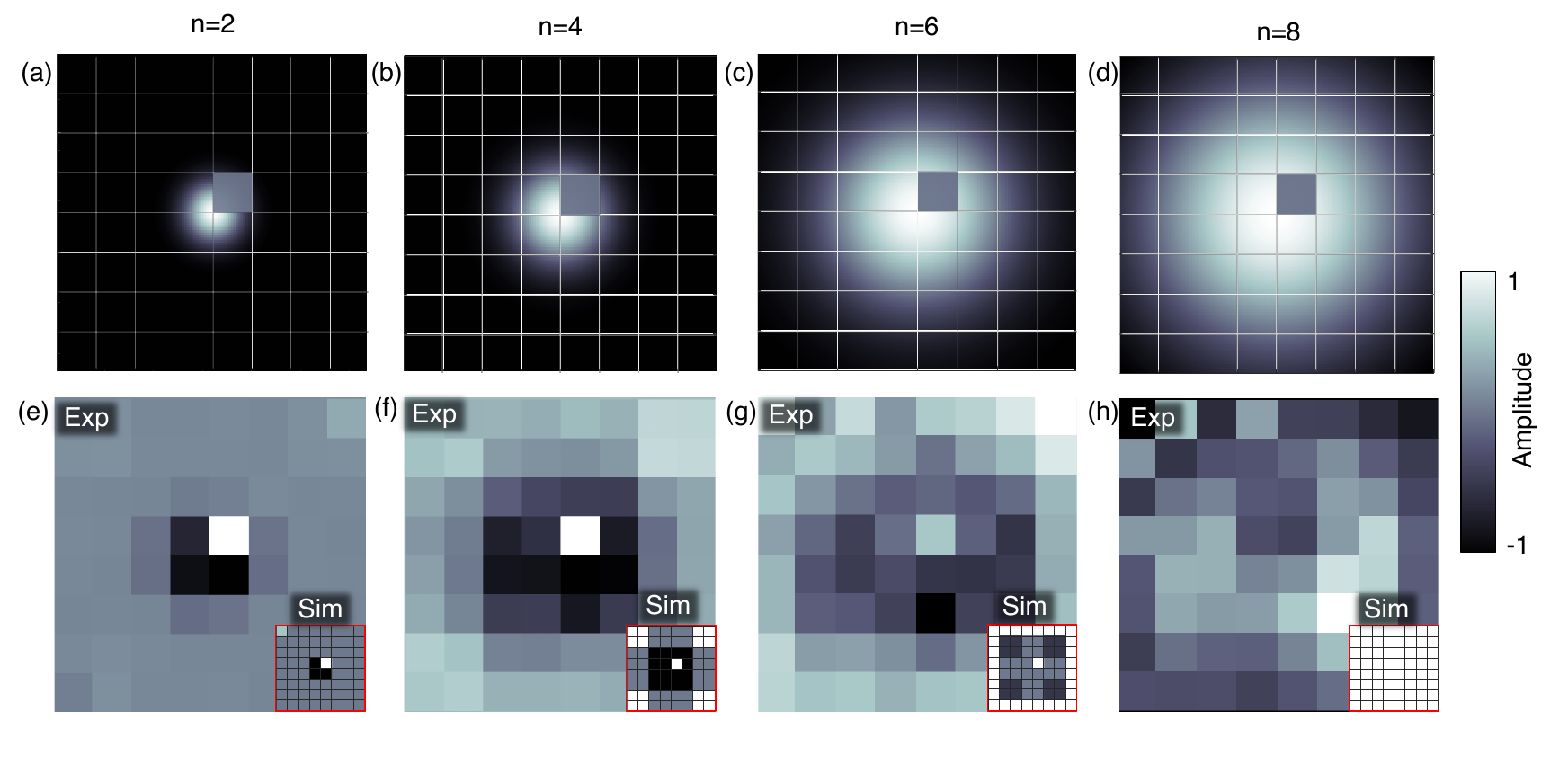}
\centering
\caption{ \textbf{Unveiling the marked elements using GI.} (\textbf{a})-(\textbf{d}) Illustrates different database sizes ($n=2,4,6,8$), respectively, relative to the Hilbert space ($m=8$). (\textbf{e})-(\textbf{h}) Demonstrates experimental results for the varying database sizes ($n=2,4,6,8$), respectively, relative to the constant Hilbert space size. The simulations for the respective values of $n$ are placed as insets.}
		\label{fig:database_exp}
\end{figure*}

\begin{figure*}[h!]
\includegraphics[width=\linewidth]{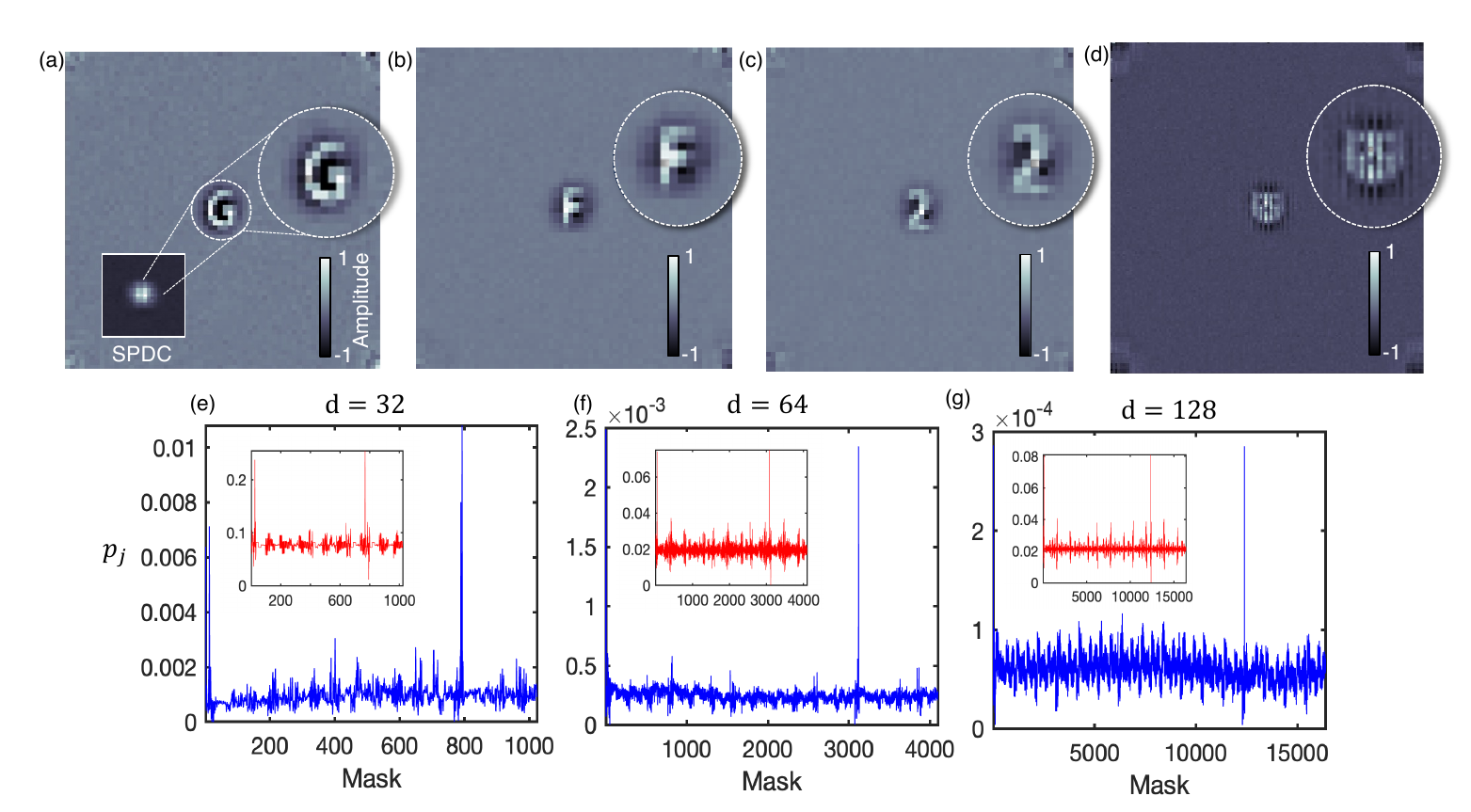}
\centering
\caption{ \textbf{Unveiling the marked elements using GI.}    (\textbf{a})-(\textbf{d}) Illustrates experimental GI results for various objects/oracles, the insets provide a zoomed in version of the results. (\textbf{a}) also provides an experimental reconstruction of the SPDC profile, which overfills the object as indicated. The 'G' object was reconstructed in different $d$-dimensional spaces, plots representing the weighting coefficients ($p_j$) for each mask in a $d$-dimensional space are shown. (\textbf{e})-(\textbf{g}) illustrates the plots for $d=32,64,128$, respectively. The insets show the theoretical predictions in each case.}
		\label{fig:examples}
\end{figure*}

Experimental results were obtained for varying relative sizes between the Hilbert space ($M = m \times m $) and database ($N=n \times n$), these relative sizes between $n$ and $m$ are illustrated in Figure \ref{fig:database_exp} (a)-(d).
In the case where $n=2$ as shown in Figure \ref{fig:database_exp} (e), there is no overlap (zero overlap) between the solution space and the non-solution space. For $n=4$ (Figure \ref{fig:database_exp} (f)) there is still no overlap between the solution and non-solution spaces since $n\leq m/2$. As $n$ increases for the cases in Figure \ref{fig:database_exp} (g) and (h) $n> m/2$, an increase in the overlap between the solution and non-solution spaces can be observed, leading to a decreasing contrast between the marked and unmarked elements. The results are in agreement with the simulations (Figure \ref{fig:solution_nonsoliution} (c), (d), (e) and (f)), showing that the overlap of the solution space and non-solution space is dependent on the size of the database (of dimensions $n$) relative to the full $m$-dimensional Hilbert space.
In this demonstration, the size of the SPDC was fixed, whereas the physical size of the marked element and mask size were varied on the SLM. The resolution of the system could also play a role if the size of the single pixel and mask resolution is higher than the system resolution. 

\vspace{0.25in}

Exemplary experimental results are shown in Figure \ref{fig:examples} (a)-(d), showing that the scheme works for various types of objects, in particular allowing for more than a single element marking and detection (analogous to GSA which can be performed on superposition states with multiple target states). In each figure, three features persist; (i) the high contrast marked elements having positive amplitudes ; (ii) the inverted Gaussian amplitude corresponding to the unmarked database elements with weightings $\lambda_j$ determined by the SPDC joint probability amplitude; (iii) pixels outside the boundary marked by $\lambda_j$. 
These features emerge theoretically from Equation \ref{eq:imreconhad}.

\vspace{0.25in}

The 'G' object/oracle was reconstructed for three different dimensions (i.e d$=32$, d$=64$, d$=128$). Plots are illustrated in Figure \ref{fig:examples} (e), (f) and (g) which represent the contributions (weighting coefficients) $p_j$ for each of the $d^2$ projective masks that were used to obtain an image reconstruction. The results for the respective dimensions are in good agreement with the theoretical contributions (which are displayed as insets in the plots). This shows that the analogy persists even through the scaling of the resolution of the object as it would be equivalent to altering the number of database elements within the fixed spatial region. 

\vspace{0.25in}

Here, we show that we do not need a diffusion operator for amplitude amplification, but rather use a measurement basis that maps the phases to amplitudes. Our results indicate that GI offers a computational advantage compared to multi-qubit search algorithms. 
We theoretically validate this by illustrating the parallels between both protocols and demonstrate how GI can be adapted to perform the same tasks. 

\section*{Conclusion}
In this work, we introduced a new perspective of GI that recasts it as a searching algorithm, reminiscent of GSA. We observed that the following statements are fulfilled:
i) Searching for an object in a distant idler photon using the position basis is similar to searching for elements in a database, where ideally, only the marked amplitudes should be amplified and detected by the process.
ii) The oracle and object are similar because they convey information about the element to be searched for in GSA and GI respectively.
iii) The search workspace constitutes the subsystem from which we perform measurements to obtain the desired elements.
Our proposal exploited this equivalence theoretically and experimentally. Using our approach, we have shown that measuring in the Grover's projection basis, i.e. $\hat{D} \ket{j}$ , it is possible to determine the same reconstruction probabilities for both protocols and a separation of the solution and non-solution subspaces is found when using traditional GI Hadamard  superpositions. This is further supported by our experimental results, for different objects and can be extended to higher dimensional spatial domains in future studies. 
%
\medskip

\textbf{Acknowledgements} \par 
N.G. would like to acknowledge the financial support received from the National Research Foundation and the National Institute for Theoretical and Computational Sciences. F.N. would like to acknowledge the financial support received from the CSIR under the HCD-IBS scholarship scheme. All authors acknowledge funding from SA QuTI.
\medskip

\textbf{Conflict of Interest} \par
The authors declare no conflict of interest.
\medskip 

\textbf{Data Availability Statement} \par
The data that supports the finding of this study are available from the corresponding author upon reasonable request.

\end{document}